\documentclass[twocolumns,structabstract]{aa}
\usepackage{amsmath}
\usepackage{graphicx}
\usepackage{txfonts}
\usepackage{natbib}
\usepackage{mathtools}
\usepackage{color}

\bibpunct{(}{)}{;}{a}{}{,} 

\begin{document}

\title{Stellar wind interaction and pick-up ion escape of the\\ Kepler-11 ``super-Earths''}

   \author{K.~G.~Kislyakova
          \inst{1}
          \and
          C.~P.~Johnstone
          \inst{2}
          \and
          P.~Odert
          \inst{1,3}
          \and
          N.~V.~Erkaev
          \inst{4,5}
          \and
          H.~Lammer
          \inst{1}
          \and
          T.~L\"{u}ftinger
          \inst{2}
          \and
          M.~Holmstr\"{o}m
          \inst{6}
          \and
          M.~L.~Khodachenko
          \inst{1,7}
          \and
          M.~G\"{u}del
          \inst{2}}


   \institute{Space Research Institute, Austrian Academy of Sciences, Schmiedlstrasse 6, A-8042 Graz, Austria
         \and
            University of Vienna, Department of Astrophysics,  T\"{u}rkenschanzstraße 17, A-1180 Wien, Austria
         \and
            Institute of Physics, University of Graz, Universit\"{a}tsplatz 5, A-8010 Graz, Austria
         \and
            Institute of Computational Modelling, Siberian Division of Russian Academy of Sciences, 660036 Krasnoyarsk, Russian Federation
         \and
            Siberian Federal University, Krasnoyarsk, Russian Federation
         \and
            Swedish Institute of Space Physics, Box 812, SE-98128 Kiruna, Sweden
         \and
            Institute of Nuclear Physics, Moscow State University, Leninskie Gory, 119992, Moscow, Russia}

   \date{Received \today}

\abstract{}
{We study the interactions between stellar wind and the extended hydrogen-dominated upper atmospheres of planets. We estimate the resulting escape of planetary pick-up ions from the five ``super-Earths'' in the compact Kepler-11 system and compare the escape rates with the efficiency of the thermal escape of neutral hydrogen atoms.}
{Assuming the stellar wind of Kepler-11 is similar to the solar wind, we use a polytropic 1D hydrodynamic wind model to estimate the wind properties at the planetary orbits. We apply a direct simulation Monte Carlo model to model the hydrogen coronae and the stellar wind plasma interaction around Kepler-11b--f within a realistic expected heating efficiency range of 15--40\%. The same model is used to estimate the ion pick-up escape from the XUV heated and hydrodynamically extended upper atmospheres of Kepler-11b--f. From the interaction model, we study the influence of possible magnetic moments, calculate the charge exchange and photoionization production rates of planetary ions, and estimate the loss rates of pick-up H$^+$ ions for all five planets. We compare the results between the five ``super-Earths'' and the thermal escape rates of the neutral planetary hydrogen atoms.}
{Our results show that a huge neutral hydrogen corona is formed around the planet for all Kepler-11b--f exoplanets. The non-symmetric form of the corona changes from planet to planet and is defined mostly by radiation pressure and gravitational effects. Non-thermal escape rates of pick-up ionized hydrogen atoms for Kepler-11 ``super-Earths'' vary between $\sim 6.4 \times 10^{30}$~s$^{-1}$ and $\sim 4.1 \times 10^{31}$~s$^{-1}$, depending on the planet's orbital location and assumed heating efficiency. These values correspond to non-thermal mass loss rates of $\sim1.07\times 10^{7}$~g$\cdot$s$^{-1}$ and $\sim6.8\times 10^{7}$~g$\cdot$s$^{-1}$ respectively, which is a few percent of the thermal escape rates.}{}

\keywords{Planet-star interactions -- Planets and satellites: atmospheres -- Planets and satellites: individual: Kepler-11 system -- Methods: numerical}

\titlerunning{Kepler-11 ``super-Earths''}
\authorrunning{K.G.~Kislyakova et al.}
\maketitle

\section{Introduction}
Due to progress in ground-based radial velocity surveys and the two space observatories, CoRoT and Kepler, an increasing number of transiting exoplanets in the ``super-Earth'' domain with radii between $\sim$1.5--3$R_{\rm \oplus}$ and masses that are $\leq 10M_{\rm \oplus}$ have been discovered. If one compares the average densities of these exoplanets to that of the Earth, one finds that the size of several ``super-Earths'' is too large to have Earth-type compositions. Only ``super-Earths'' orbiting their host stars at distances below 0.05 AU, such as CoRoT-7b \citep{Leger2009}, Kepler-10b \citep{Batalha2011}, and Kepler-18b \citep{Cochran2011}, show radius-mass related average densities that agree with pure rocky bodies without gaseous envelopes. This radius-mass anomaly indicates that the majority of each of these exoplanets contains a huge fraction of their primordial nebula-based hydrogen-dominated protoatmospheres or volatile materials, such as evaporated ices and hydrogen-rich H$_2$O, CH$_4$, and NH$_3$. The result that such volatile-rich exoplanets should exist in the inner regions of exosolar planetary systems was predicted nearly a decade ago by \citet{Kuchner2003} and \citet{Leger2004}. In this context, it is interesting that the compact ``super-Earths'' of the Kepler-11 system, Kepler-11d, Kepler-11e, and Kepler-11f, are most likely to contain large volumes of hydrogen, while Kepler-11b and Kepler-11c may be rich in ices similar to Uranus and Neptune surrounded by a H/He mixture \citep{Lissauer2011}.

\begin{table*}
\renewcommand{\baselinestretch}{1}
\caption{Planetary and stellar parameters for the Kepler-11 system given in Earth masses and radii. Values in [m] and [kg] were used in the simulations (\citealp{Lissauer2011}, http://kepler.nasa.gov/Mission/discoveries/, http://www.exoplanet.eu).}
\begin{center}
\begin{tabular}{ccccccc}
\hline\hline
Exoplanet/star  & $d$ [AU] & $i$, [deg] & $R_{\rm pl}$ [$R_{\rm \bigoplus}$]  & $R_{\rm pl}$ [m] & $M_{\rm pl}$ [$M_{\rm \bigoplus}$] & $M_{\rm pl}$ [kg]  \\\hline
Kepler-11  &   613 [pc] &  --  & 1.1$\pm$0.1 [$R_{\rm \bigodot}$] & $7.65\times 10^8$ & 0.95$\pm$0.1 [$M_{\rm \bigodot}$] & $1.89 \times 10^{30}$ \\
Kepler-11b &   0.091    & 88.5 & 1.97$\pm$0.19 & $1.2 \times 10^7$  & 4.3$_{-2.0}^{2.02}$ & $2.57 \times 10^{25}$ \\
Kepler-11c &   0.1      & 89.0 & 3.15$\pm$0.3  & $2.0 \times 10^7$  & 13.5$_{-6.1}^{4.8}$ & $8.06 \times 10^{25}$ \\
Kepler-11d &   0.159    & 89.3 & 3.43$\pm$0.32 & $2.18 \times 10^7$ & 6.1$_{-1.7}^{3.1}$  & $3.64 \times 10^{25}$ \\
Kepler-11e &   0.195    & 88.8 & 4.52$\pm$0.43 & $2.88 \times 10^7$ & 8.4$_{-1.9}^{2.5}$  & $5.01 \times 10^{25}$ \\
Kepler-11f &   0.25     & 88.4 & 2.61$\pm$0.25 & $1.66 \times 10^7$ & 2.3$_{-1.2}^{2.2}$  & $1.37 \times 10^{25}$ \\
\hline
\end{tabular}
\end{center}
\normalsize
\label{t_1}
\end{table*}

Recently, \citet{Lammer2013} studied the non-hydrostatic upper atmosphere structure and blow-off criteria of seven ``super-Earths'' with known size and masses, which included Kepler-11b--f. By assuming that the thermospheres of these planets are dominated by hydrogen, the results of this study indicate that their upper atmospheres may expand to several planetary radii and, with the exception of Kepler-11c, have exobase levels that extend beyond the Roche lobe, indicating that they all experience atmospheric mass-loss due to Roche lobe overflow (e.g., \citealp{Lecavelier2004, Erkaev2007}). \citet{Lammer2013} found that the atmospheric mass loss of the studied ``super-Earths'' is one to two orders of magnitude lower compared to that of ``hot Jupiters'', such as HD 209458b, so that one can expect that these exoplanets cannot lose their present hydrogen envelopes via thermal escape during their remaining lifetimes.

Because the upper atmospheres of these planets can expand to large distances, one can assume that huge extended hydrogen coronae should form around the planets above a possible magnetopause obstacle, so that planetary atoms can directly interact with the stellar wind plasma. This results in non-thermal ion pick-up escape.

The aim of this study is to investigate the expected hydrogen coronae around these five ``super-Earths'' within the Kepler-11 system, the stellar wind, and XUV induced H$^+$ pick-up ion escape rates and to compare their efficiency with the thermal loss rates given in \citet{Lammer2013}. For the first time this study allows us to obtain an estimate of the thermal and non-thermal ion pick-up escape rates from observed ``super-Earths'', which are located in different orbital distances inside the same planetary system. Table~\ref{t_1} summarizes the parameters of the Kepler-11 system. For Kepler-11, the distance between this star and the Sun is given. In the present study, we do not consider Kepler-11g which is a Jupiter-type gas giant.

We apply a direct simulation Monte Carlo (DSMC) exosphere-stellar wind plasma interaction model \citep{Holmstrom2008, Ekenback2010, Kislyakova2013} to the soft X-ray and extreme ultraviolet (XUV) heated and dynamically expanded upper atmospheres of Kepler-11b--f. In Section~\ref{sec_star}, we describe the radiation, expected stellar wind plasma parameters, and the magnetic properties of the host star. In Section~\ref{sec_modeling}, we briefly describe the DSMC model, which is used for the calculation of the hydrogen coronae and the coupled solar/stellar wind plasma model. We also briefly address possible planetary obstacles and expected magnetospheres of these planets. In Section~\ref{sec_pickup}, we calculate the ion production rates and estimate the resulting ion pick-up escape and mass loss rates. Finally, we compare the H$^+$ ion pick-up escape rates with the thermal escape rates of the neutral hydrogen atoms.

\section{Radiation and stellar wind properties}
\label{sec_star}

\subsection{Estimated XUV and Ly$\alpha$ emission of Kepler-11}
\label{ssec_K11_XUV}

\begin{table*}
\renewcommand{\baselinestretch}{1}
\caption{Stellar input parameters at the orbit locations of Kepler-11b-f.}
\begin{center}
\begin{tabular}{ccccccccc}
\hline\hline
Exoplanet & $I_{\rm XUV}$ & $\beta_{\rm abs}$ [s$^{-1}$] & $\tau_{\rm pi}$ [s$^{-1}$] & $\tau_{\rm ei}$ [s$^{-1}$] & $v_{\rm sw}$ [km s$^{-1}$] & n$_{\rm sw}$ [cm$^{-3}$] & $T_{\rm el}$ [K]     & $B_{\rm IMF}$ [G]\\\hline
Kepler-11b & 60  & 0.174 & 6.7$\times 10^{-6}$   & 3.16 $\times 10^{-5}$  & 394.0  & 860.0    & 1.25$\times 10^{6}$ & 3.8$\times 10^{-3}$ \\
Kepler-11c & 45  & 0.128 & 5.04$\times 10^{-6}$  & 1.58 $\times 10^{-5}$  & 404.0  & 635.0    & 1.0$\times 10^{6}$  & 3.0$\times 10^{-3}$ \\
Kepler-11d & 20  & 0.057 & 2.26$\times 10^{-6}$  & 3.16 $\times 10^{-6}$  & 419.0  & 245.0    & 6.3$\times 10^{5}$  & 1.2$\times 10^{-3}$ \\
Kepler-11e & 13  & 0.038 & 1.45$\times 10^{-6}$  & 1.0$\times 10^{-6}$    & 423.0  & 160.0    & 4.0$\times 10^{5}$  & 8.0$\times 10^{-4}$ \\
Kepler-11f & 8   & 0.023 & 8.96$\times 10^{-7}$  & 7.97$\times 10^{-7}$   & 428.0  & 95.0     & 2.5$\times 10^{5}$  & 5.0$\times 10^{-4}$ \\
\hline
\end{tabular}
\end{center}
\normalsize
\label{t_2}
\end{table*}

Kepler-11 is a slightly evolved, solar-like G-star with a radius $R_{\rm st}$ of 1.1$\pm$0.1$R_{\rm Sun}$ and a mass $M_{\rm st}$ of 0.95$\pm$0.1$M_{\rm Sun}$ \citep{Lissauer2011}. At present, the X-ray and EUV (XUV) emission of Kepler-11 is observationally unconstrained. This is mainly due to its large distance of several hundred parsecs and its old age (6--10~Gyr; \citealt{Lissauer2011}), which implies a low level of activity. Therefore, we adopt the XUV fluxes at the respective planetary orbits given in \citet{Lammer2013}, which are based on power laws derived from observations of nearby solar-like stars of different ages \citep{Ribas2005}. At the orbits of Kepler-11b--f, the XUV fluxes are enhanced by factors $I_{\rm XUV}$ of 60, 45, 20, 13, and 8 (cf. Table~\ref{t_2}) over the average present solar XUV flux at 1~AU ($4.64\,\mathrm{erg\,cm^{-2}\,s^{-1}}$; \citealp{Ribas2005}). Note that the uncertainties in $I_{\rm XUV}$ might be up to an order of magnitude because of the lack of relevant observations and weak constraints on the age or activity level of the host star Kepler-11 \citep{Lammer2013}. The enhancement factors are used to calculate the photoionization rates, which are needed as input for the DSMC ENA model presented here. The mean photoionization rate of $1.1\times10^{-7}\,\mathrm{s^{-1}}$ at the present Earth (e.g., \citealp{Bzowski2008}) is scaled up by $I_{\rm XUV}$ and corresponds to $\tau_{\rm pi}$ values for planets Kepler-11b--f given in Table \ref{t_2}.

Another required model input parameter is the UV absorption rate, which is defined as $\beta_\mathrm{abs}=\int \sigma(\lambda) \Phi(\lambda) \mathrm{d}\lambda$, where $\sigma(\lambda)$ is the absorption cross-section and $\Phi(\lambda)$ is the stellar Ly$\alpha$ spectrum (e.g. \citealp{Meier1995}). The absorption cross-section is $\sigma(\lambda)=\int \psi(v) \sigma_N(\lambda') \mathrm{d}v$, where $\psi(v)$ is the normalized atomic velocity distribution and $\sigma_N(\lambda')$ is the natural absorption cross-section of an atom traveling with velocity $v$ for which $\lambda'=\lambda(1-v/c)$. When the atomic velocities are very small, the UV absorption rate can be approximated by the product of the total cross-section at Ly$\alpha$, $\sigma_\mathrm{Ly\alpha}=\int \sigma(\lambda) \mathrm{d}\lambda=5.47\times10^{-15}\,\mathrm{cm^2}$\,\AA\ (e.g. \citealp{Quemerais2006}), and the stellar Ly$\alpha$ photon flux at the line center. If we estimate the central Ly$\alpha$ flux of Kepler-11 using a scaling law from \citet{Ribas2005} for the total line flux as a function of stellar age and divide it by the effective line width ($\sim$1\,\AA; \citealp{Meier1995}), we obtain the UV absorption rates $\beta_{\rm abs}$ for Kepler-11b--f given in Table~\ref{t_2}.

The UV absorption rates might deviate from the values given above, especially for atoms with high radial velocities. Therefore, we calculate a velocity-dependent UV absorption rate. For this, we need the intrinsic (unabsorbed) stellar Ly$\alpha$ line profile, which is not available for Kepler-11. Hence, we adopt the reconstructed intrinsic Ly$\alpha$ profile of the Sun-like star 61~Vir presented in \citet{Wood2005}. This G5V star is located at a distance of approximately 8.5~pc and has an estimated age range of 6--12~Gyr \citep{Vogt2010}. Despite very similar masses, 61 Vir appears to be smaller than Kepler-11, indicating that it is probably still younger. However, since it has the weakest intrinsic Ly$\alpha$ emission observed from Sun-like stars to date \citep{Linsky2013}, we adopt its profile as a proxy for to profile of Kepler-11. To obtain the Ly$\alpha$ fluxes at the orbits of the Kepler-11 planets, we scale the apparent intrinsic flux of 61~Vir \citep{Wood2005} according to the star's distance (8.555~pc; \citealp{Linsky2013}) and the orbital distances of the Kepler-11 planets (cf. Table~\ref{t_1}). The line profile has also been shifted in wavelength to account for the radial velocity of 61~Vir ($-8.1$\,km\,s$^{-1}$; \citealp{Kharchenko2007}).

\begin{figure}
\centering
\includegraphics[width=1.0\columnwidth]{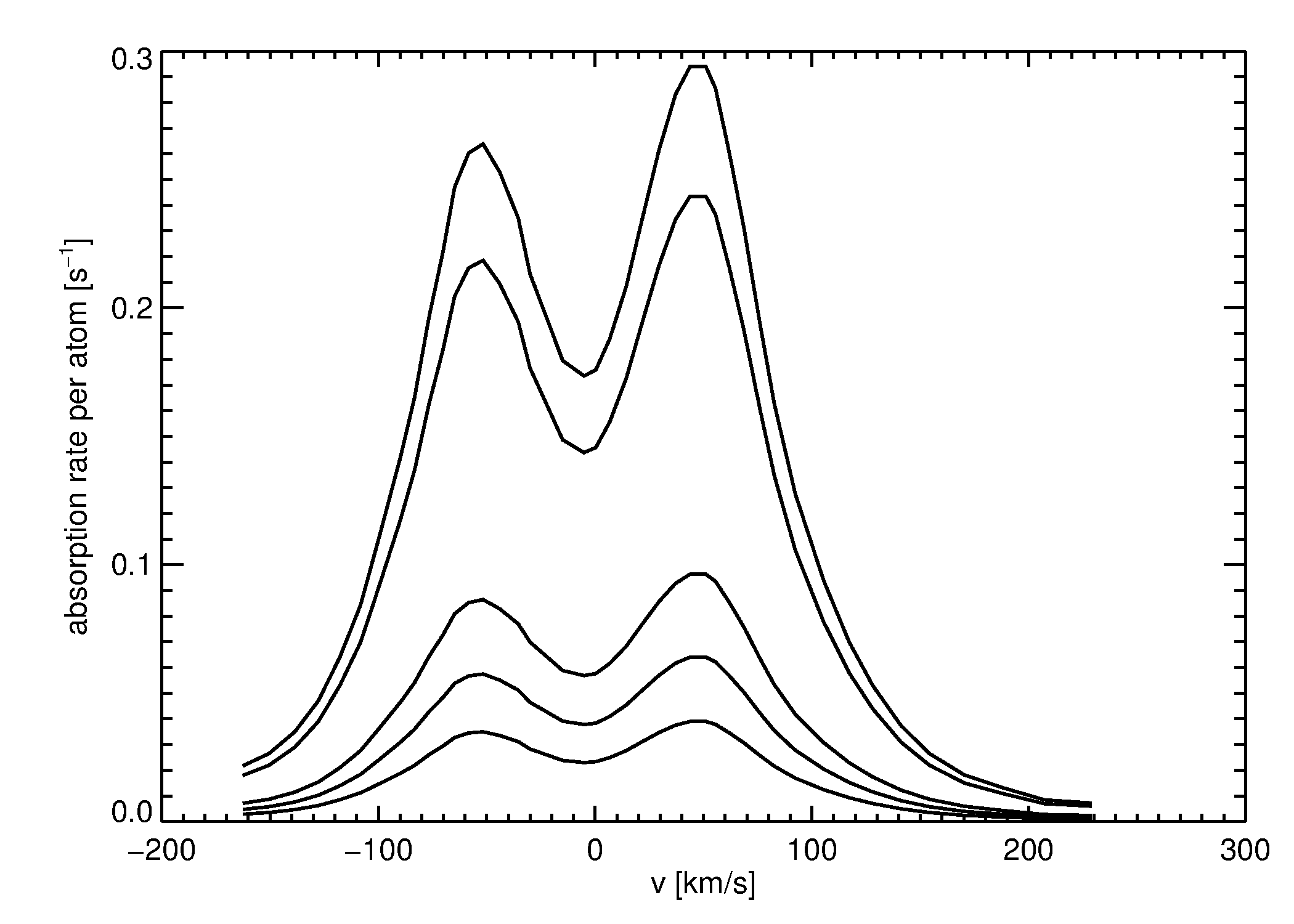}
\caption{Kepler 11 Ly$\alpha$ absorption rate as a function of neutral hydrogen radial velocity. The lines from top to bottom correspond to the orbit locations of Kepler-11b--f.}
\label{f_UVar}
\end{figure}

We define the velocity-dependent UV absorption rate per atom $\beta(v)$ such that $\beta_\mathrm{abs}=\int \psi(v) \beta(v) \mathrm{d}v$. Therefore, $\beta(v)=\int \sigma_N(\lambda') \Phi(\lambda) \mathrm{d}\lambda$, as can be derived from the equations given above. Because the natural absorption cross-section is very narrow compared to the stellar line profile, it can be approximated by a delta function; therefore, $\beta(v)\approx\Phi(\lambda_0(1-v/c))\int \sigma_N(\lambda') \mathrm{d}\lambda$, where $\lambda_0=1215.67\AA$ is the central wavelength of Ly$\alpha$. The resulting absorption rates at the orbits of Kepler-11b--f as a function of atomic radial velocity are shown in Fig.~\ref{f_UVar}. We define velocities in the direction of the star to be positive. Note that the values found in the line center, i.e. for atoms with $v\approx0$, are in good agreement with the simple calculations given in Table~\ref{t_2}. However, we use the rates shown in Fig.~\ref{f_UVar} in the DSMC simulations presented below for better accuracy.

\subsection{Expected stellar wind plasma properties}
\label{ssec_K11_sw}

In this section, we estimate stellar wind speeds, densities, temperatures, and magnetic field strengths at the planetary orbital radii. Since the properties of the winds of low-mass stars are unconstrained by both theory and observation, we assume that the stellar wind of Kepler-11 is similar to the solar wind. This is a reasonable assumption, given the similarities between Kepler-11 and the Sun.

\begin{figure}
\centering
\includegraphics[width=0.8\columnwidth]{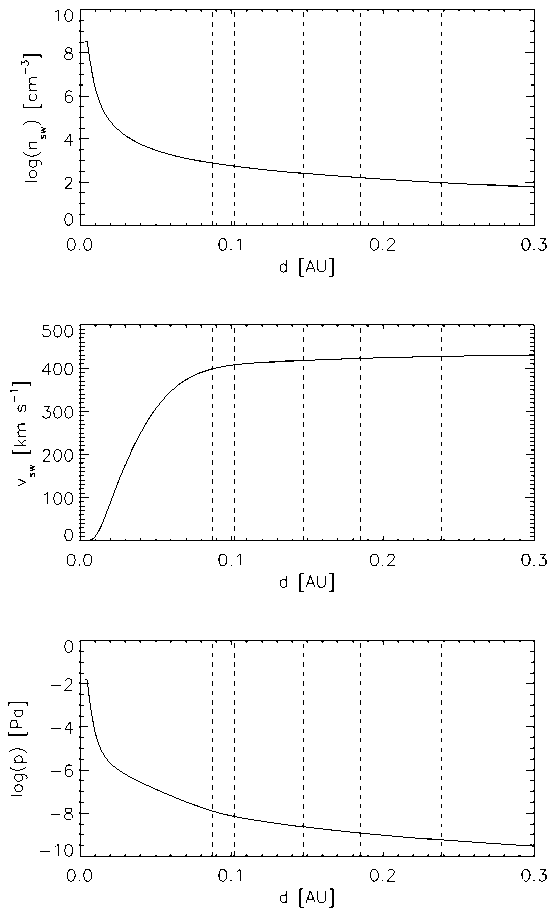}
\caption{Modeled stellar wind parameters ($n_{\rm sw}$, $v_{\rm sw}$, $p_{\rm sw}$) as a function of orbital distance.
The dashed lines correspond to the orbit locations of Kepler-11b--f.}
\label{f_wind}
\end{figure}

We apply a 1D hydrodynamic model for the solar wind propagation by using the publicly available MHD code Nirvana \citep{Ziegler2005}. We assume that the wind is driven by thermal pressure gradients. In this model, the acceleration of the wind is governed by the amount and distribution of heating that is given to the wind.
Since the heating mechanisms are unknown, we implicitly heat the wind by assuming a polytropic equation of state, where the pressure is assumed to be proportional to the mass density to the power of a polytropic index $\alpha$, thus giving $p = K \rho^{\alpha}$ (this is equivalent to assuming that the temperature is give by $T_{sw} \propto \rho^{\gamma-1}$). This assumption means that we do not solve the energy equation in our hydrodynamical simulations.  Assuming radial variations in the values of $K$ and $\alpha$, \citet{Jacobs2011} produced a series of 1D models for the solar wind that reproduces a range of slow and fast solar wind properties at 1AU as measured by the ACE satellite. We reproduce one of their models that gives typical slow solar wind properties at 1~AU. Given that the planets are likely to orbit close to the stellar equatorial plane (cf. Table~\ref{t_1}) by analogy with the Sun, it is reasonable to assume that they experience mostly slow solar wind conditions, which dominate in the equatorial plane throughout most of the solar cycle for the solar wind \citep{Ebert2009}. In this model, we assume a particle number density at the base of the wind of $7.0 \times 10^{8}$ cm$^{-3}$ and a temperature of $1.625 \times 10^{6}$ K. The $\alpha$ parameter varies smoothly from a value of 1.22 near the stellar surface to 1.42 at 23 R$_{\rm st}$ and is then uniform out to 1AU. This leads to a wind proton density of 5.3 protons cm$^{-3}$ and a temperature of $1.3 \times 10^5$ K at 1~AU. These are reasonable values for the slow solar wind, as measured by the ACE satellite (e.g. \citealp{Wargelin2004, Vrsnak2007, Manoharan2012}). The mass density, velocity, and pressure of the wind as a function of radial distance from the star are shown in Fig. \ref{f_wind}.

To estimate the strength of the stellar magnetic field far from the star, it is necessary to know the strength of the dipole component of the field close to the star. Since there exists no measurements of the strength of Kepler-11's magnetic field, we scale the strength of the solar dipole to Kepler-11 using the rotation-activity relation given by \citet{Saar1996} of $\bar{B} \propto P_{\rm rot}^{-1.7}$. We estimate the rotation period of Kepler-11 using its age and the rotation-age relation of $\log(1/P_{\rm rot}) = 1.865 - 2.254 t^{0.083}$ given by \citet{Guinan2009}, where $P_{\rm rot}$ is the rotation period in days and $t$ is the stellar age in Gyr. This gives a rotation period for Kepler-11 of 33.5 days, and assuming that the inclination angle of the axis-of-rotation is 90$^\circ$, this corresponds to a projected rotational velocity at the equator of 1.66 km s$^{-1}$. We check that this value is reasonable by computing an atmospheric model using {\sc LLmodels} \citep{Shulyak2004} and calculating synthetic spectra with synth3 \citep{Kochukhov2007}, which we then compare to spectroscopic observations obtained with the Keck I telescope and de-archived from KOA (Keck Observatory Archive). The basic physical parameters of the star, $T_{\rm eff}$= 5680 K $\pm$ 100\,K and $\log g$=4.3 $\pm$ 0.2 are adopted from \citet{Lissauer2013}. The atomic data used for spectrum synthesis is extracted from the Vienna Atomic Line Database (VALD, \citealp{Piskunov1995, Ryabchikova1999, Kupka1999}). Comparing the synthetic spectra to selected suitable spectral lines from the Keck I observations, we get an estimate for the projected rotational velocity of Kepler-11 of 1.5 $\pm$ 1 {km\,s $^{-1}$}, which is similar to the value calculated above. Assuming a dipole field strength of 5 G for the Sun, we get a dipole field strength of 3.5 G for Kepler-11.

For the radial dependence of the strength of the dipole component, we assume that the field decreases as a dipole (i.e. $B \propto r^{-3}$) out to the source-surface, where the coronal plasma pressure dominates the magnetic pressure and forces the magnetic field lines open. Outside of the source-surface, we assume that the field falls off as a radial field (i.e. $B \propto r^{-2}$). Based on the solar analogy, we take the source-surface to be at 2.5 R$_{\rm st}$. This model gives a value of 4.3 nT at 1~AU for the solar case, and the field strengths at each planetary orbital distance for Kepler-11 planets are given in Table~\ref{t_2}.

\section{Upper atmosphere and stellar wind interaction modeling}
\label{sec_modeling}

\subsection{Method}

We study the plasma interactions between the stellar wind of Kepler-11 and the upper atmospheres of the Kepler-11 ``super-Earths'' by applying a direct simulation Monte Carlo (DSMC) upper atmosphere-exosphere 3D particle model. The software is based on the FLASH code written in Fortran 90, which is publicly available \citep{Fryxell2000}. The code follows particles in the simulation domain according to all forces acting on a hydrogen atom and includes two species: neutral hydrogen atoms and hydrogen ions (which also include stellar wind protons). The application of the FLASH code to the physics of upper planetary atmospheres is described in detail in \citet{Holmstrom2008}, \citet{Ekenback2010}, and \citet{Kislyakova2013}. The model includes the following processes/forces, which may act on an exospheric atom:
\begin{enumerate}
  \item Collision with an UV photon, which can occur if the particle is outside of the planet's shadow, which leads to an acceleration of the hydrogen atom away from the star. As in \citet{Holmstrom2008, Ekenback2010} a UV photon is absorbed by a neutral hydrogen atom, leading to a radial velocity change and then consequently reradiated in a random direction. However, in the present study, the UV collision rate is velocity dependent, as shown in Fig.~\ref{f_UVar}.
  \item Photoionization by a stellar photon or a stellar wind electron impact ionization. The rates, $\tau_{\rm pi}$ and $\tau_{\rm ei}$, are presented in Table~\ref{t_2}, and $\tau_{\rm ei}$  is taken according to \citet{Holzer1977}. Estimation of the photoionization rate $\tau_{\rm pi}$ is described in Section \ref{ssec_K11_XUV}. We do not include ionization by the atmospheric electrons, because electron ionization rate strongly depends on the electron temperature and is very low for the atmospheric values of $T_{\bf el}$ \citep{Holzer1977}.
  \item Charge exchange between neutral hydrogen atoms and stellar wind protons. If a hydrogen atom is outside the planetary obstacle (magnetopause, or ionopause) it can undergo charge exchange with a stellar wind proton, producing an energetic neutral atom (ENA). The charge exchange cross-section is taken to be equal to $2 \times 10^{-19}$~m$^2$ \citep{Lindsay2005}.
  \item Elastic collision with another hydrogen atom. Here the collision cross-section was taken to be $10^{-21}$~m$^2$ \citep{Izmodenov2000}.
\end{enumerate}

Collisions with cold H$^+$ of  planetary origin are not taken into account. These ions are abundant in particular at the border of the planetary obstacle, but they should not alter the results significantly. Although the cross-section is large for low velocity H-H$^+$ \citep{Lindsay2005}, neutral H dominates at low energies by orders of magnitudes (in our model), so the effect of  charge exchange should be minimal at low energies.

The coordinate system is centered at the center of the planet with the $x$-axis pointing towards the center of mass of the system, the $y$-axis pointing in the direction opposite to the planetary motion, and the $z$-axis pointing parallel to the vector $\boldsymbol{\Omega}$ that represents the orbital angular velocity of the planet around the central star. The $M_{\rm st}$ is the mass of the planet's host star. The outer boundary of the simulation domain is the box
$x_{\rm min} \leq x \leq x_{\rm max}$,~
$y_{\rm min} \leq y \leq y_{\rm max}$, and
$z_{\rm min} \leq z \leq z_{\rm max}$. The inner boundary is a sphere of radius $R_0$.

We obtain the planetary input parameters, such as density, temperature, and bulk outflow velocity of the upper atmosphere, for the inner boundary in the 3D DSMC particle code in the same way as described in detail in \citet{Lammer2013}. Our assumed lower thermospheric conditions are also similar to those assumed in \citet{Lammer2013}. We used the time-dependent numerical algorithm which solves the 1D fluid equations for mass, momentum, and energy conservation from the base of the thermosphere until the height $R_0$ where the Knudsen number $Kn$ is 0.1 \citep{Johnson2013}. The Knudsen number is the ratio of the mean free path to the corresponding scale height, and the exobase is defined as the surface where $Kn = 1$. Using kinetic 3D modeling above $R_0$ is more accurate because, as we show, the hydrogen coronae are strongly deformed by the radiation pressure for the Kepler-11 planets.

Tidal potential, Coriolis force and centrifugal force, and the gravitation of the star and planet, which act on a hydrogen neutral atom are included in the following way (after \citealp{Chandrasekhar1963}):
\begin{multline}
  \frac{d v_{\rm i}}{dt}=\frac{\partial}{\partial x_{\rm i}} \Bigg[ \frac{1}{2}\Omega^2 \left(x_1^2+x_2^2 \right)+ \mu\left(x_1^2-\frac{1}{2}x_2^2 -\frac{1}{2}x_3^2 \right)+\\
+\left(\frac{G M_{\rm st}}{R^2} - \frac{M_{\rm st} R}{M_{\rm pl} + M_{\rm st}}\Omega^2 \right)x_1 \Bigg] + 2\Omega \epsilon_{\rm il3} v_{\rm l}
  \label{e_SCG}
\end{multline}
Here $x_1 = x$, $x_2 = y$, $x_3 = z$, and $v_{\rm i}$ are the components of the velocity vector of a particle; $G$ is Newton's gravitational constant; $R$ is the distance between the centers of mass; $\epsilon_{\rm ilk}$ is the Levi-Civita symbol; and $\mu$ is given by $\mu = G M_{\rm st}/R^3$. The first term in the right-hand side of Equation \ref{e_SCG} represents the centrifugal force, and the second is the tidal-generating potential; the third corresponds to the gravitation of the planet's host star and the planet, while the last term stands for the Coriolis force. The self-gravitational potential of the particles is neglected.

\subsection{Possible planetary obstacles}

Charge exchange reactions between stellar wind protons and neutral planetary particles consist of the transfer of an electron from a neutral atom to a proton that leaves a cold atmospheric ion and an energetic neutral atom. This process is described by the following reaction: $H_{\rm sw}^+ + H_{\rm pl} \to H^+_{\rm pl} + H_{\rm ENA}$. After its production, an ENA continues to travel with the initial velocity and energy of the stellar wind proton. The atmospheric atom becomes an initially cold ion, which can afterwards be lost from the atmosphere due to the ion pick-up process (see Section \ref{sec_pickup}).

Charge exchange takes place outside of an obstacle that corresponds to the magnetopause or ionopause boundary, which we assume is a surface described by:
\begin{equation}
	x = R_{\rm s} \left(1 - \frac{y^2 + z^2}{R_{\rm t}^2} \right),
\label{e_obs}
\end{equation}

\noindent where $R_{\rm s}$ is the substellar point of the obstacle and $R_t$ is the obstacle width at the planetary terminator. Changing these two parameters, one may model various types of the interaction, such as the interaction between the stellar wind and a magnetized or non-magnetized body. More details are given in \citet{Kislyakova2013}. The obstacle is rotated by an angle of $\arctan (v_{\rm p}/v_{\rm sw})$ to account for the finite stellar wind speed relative to the planet's orbital speed. The protons do not penetrate a magnetic obstacle.

For magnetized planets, the dynamical pressure of the stellar wind is balanced by the pressure of the atmosphere together with the magnetic pressure defined by the intrinsic magnetic moment of the planet. Types of expected magnetospheres of exoplanets were considered in detail, for example, by \citet{Khodachenko2012}. Because the magnetic fields of the Kepler-11 planets are unknown, we focus on non-magnetized planets without additional magnetic protection by intrinsic dynamos or magnetic discs in the present study \citep{Khodachenko2012}.
Under this assumption, the dynamic pressure of the Kepler-11 wind can be balanced by the weak atmospheric pressure in the upper layers of the atmospheres of Kepler-11b--f only very close to $R_0$. For simplicity, we do not consider the magnetic pressure from induced currents in the ionosphere. Since the hydrogen coronae around the planets are not symmetric (see the following section), this moves the interaction boundary close to the planet. In our simulations, we assume $R_{\rm s}$ to be slightly above $R_0$ and $R_{\rm s} = R_{\rm t}$, which corresponds to a Venus-type narrow magnetic obstacle. Magnetically non-protected planets undergo the strongest interaction processes. Thus, this approach allows us to estimate the highest possible pick-up escape rates.

\subsection{Hydrogen coronae and modeling results}

\begin{table*}
\renewcommand{\baselinestretch}{1}
\caption{Planetary and upper atmosphere input parameters at the height $R_{0~\rm \eta=15\%,\eta=40\%}$ where the $Kn=0.1$, as the starting point for the DSMC-code simulations for two heating efficiencies $\eta=$15\% and 40\%.}
\begin{center}
\begin{tabular}{ccccccccc}
\hline\hline
Exoplanet & $R_{\rm0~\eta=15\%}$ [$R_{\rm pl}$] & $n_{\rm \eta=15\%}$ [cm$^{-3}$] & $T_{\rm \eta=15\%}$ [K] & $v_{\rm flow~\eta=15\%}$ [km/s] & $R_{\rm 0~\eta=40\%}$ [$R_{\rm pl}$]  & $n_{\rm \eta=40\%}$ [cm$^{-3}$] & $T_{\rm \eta=40\%}$ [K] & $v_{\rm flow~\eta=40\%}$ [km/s]\\\hline
Kepler-11b & 6.7 & $7 \times 10^{5}$  & 1440 &  2.3  & 6.7  &  $8.2 \times 10^{5}$  & 2230 &  4.0  \\
Kepler-11c & 5.0 & $1 \times 10^{6}$  & 1833 &  0.78 & 6.5  &  $4.8 \times 10^{5}$  & 2582 &  1.7  \\
Kepler-11d & 6.4 & $5 \times 10^{5}$  & 1000 &  0.96 & 7.0  &  $5.5 \times 10^{5}$  & 1453 &  2.1  \\
Kepler-11e & 6.5 & $4 \times 10^{5}$  & 770  &  0.75 & 7.5  &  $3.78 \times 10^{5}$ & 1234 &  1.7  \\
Kepler-11f & 18  & $3.2\times 10^{5}$ & 761  &  1.6  & 22   &  $2.0 \times 10^{5}$  & 397  &  2.8  \\
\hline
\end{tabular}
\end{center}
\normalsize
\label{t_3}
\end{table*}

\begin{figure*}
\centering
\includegraphics[width=2.0\columnwidth]{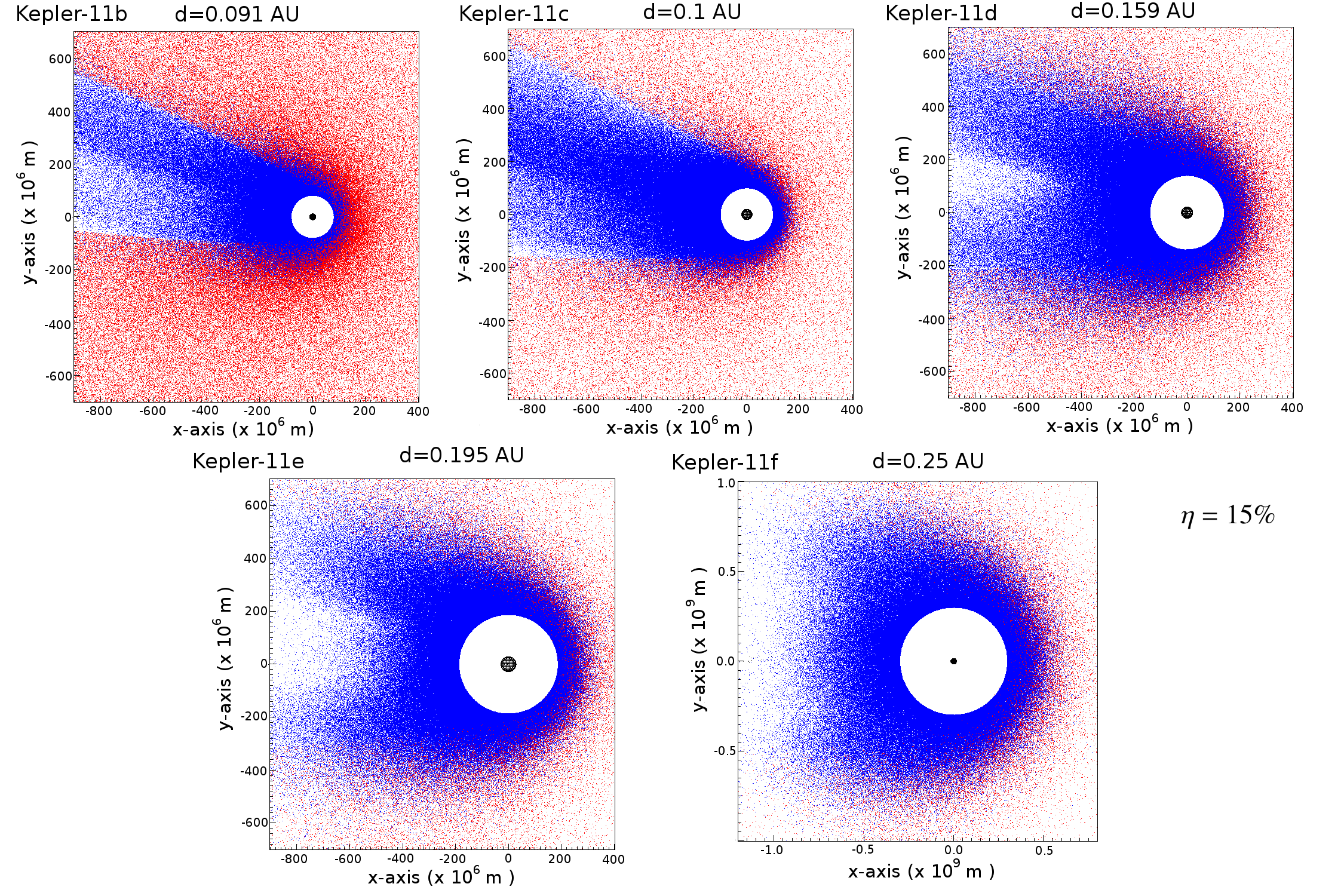}
\caption{Slices of modeled 3D atomic hydrogen coronae around the five Kepler-11 ``super-Earths'' for  $-10^7 \le z \le 10^7$~m with a heating efficiency $\eta$ = 15\%. Blue and red dots correspond to neutral hydrogen atoms and hydrogen ions, which include stellar wind protons, respectively. The black dot in the center represents the planet. The white empty area around the planet corresponds to the XUV heated, hydrodynamically expanding thermosphere up to the height $R_0$, where $Kn$=0.1.}
\label{f_clouds}
\end{figure*}

In this section, we present the results of the modeled hydrogen clouds around the five Kepler-11 ``super-Earths''. Huge coronae of neutral hydrogen are formed around each of these low density ``super-Earths''. Figure~\ref{f_clouds} illustrates the appearance of the hydrogen coronae near the planets for a heating efficiency, $\eta$, of 15\%. The heating efficiency corresponds to the ratio of the absorbed XUV energy that is transferred into heat of the neutral gas. Values of 15\% and 40\% are considered to be the most realistic (see discussion in \citealp{Chassefiere1996, Erkaev2013, Lammer2013}). The atmospheric input parameters are given in Table~\ref{t_3}. In Table~\ref{t_3}, $R_0$ corresponds to the simulation starting level given in planetary radii, and $n$ and $T$ stand for the atmospheric density and temperature at the same height. Finally, $v_{\rm flow}$ denotes the bulk outflow velocity at $R_0$, which is caused by the strong XUV heating of the atmosphere and the resulting hydrodynamic expansion (see \citealp{Lammer2013}). All parameters are given for both $\eta$=15 \% and 40\%. Stellar wind and radiation input parameters are described above in Section \ref{sec_star} and can be found in Tables \ref{t_1} and \ref{t_2} together with planetary masses and radii.

As one can see from Fig.~\ref{f_clouds}, the hydrogen coronae around the Kepler-11 planets strongly deviate from the spherically symmetric form assumed by \citet{Lammer2013}, where radiation pressure effects were not taken into account. There are several reasons for the modifications of the shapes of the coronae: first, the radiation pressure from Kepler-11 accelerates the neutral atoms that are outside the optical shadow of the planet away from the star, which leads to the formation of a cometary-like tail; second, charge-exchange and electron impact ionization reduce the neutral hydrogen density outside the magnetic obstacle. Third, the tidal force leads to the deformation (twist) of the cloud. Although these effects for these planets are not so strong in comparison to ``Hot Jupiters'', such as HD 209458b or HD189733b \citep{Bourrier2013}, they are of significant importance. Unlike electron impact ionization, photoionization can also occur inside the planetary obstacle but only outside the optical shadow. The hydrogen atoms that have undergone charge-exchange are identified by an enhanced H$^+$ density near the interaction border along the planetary obstacle.

The high radiation pressure of Kepler-11 with gravitational effects, leads to the formation of non-symmetric tails on both sides of a planet in any given plane. In 3D, these tails correspond to a non-symmetric ring of accelerated neutral hydrogen atoms. For a heating efficiency of 40\%, the atmospheres are slightly more inflated, but the coronae do not differ much qualitatively from those shown in Fig. \ref{f_clouds}.
One can also see that planets in orbital locations further from their host star, where either radiation pressure or tidal forces are weaker, have hydrogen coronae that differ only slightly from spherical forms \citep{Kislyakova2013}. However, these effects cannot be neglected in the case of close-in exoplanets.

It should be mentioned that the so-called self-shielding effect in some cases can significantly diminish the acceleration by radiation pressure, photoionization of neutral hydrogen atoms \citep{Bourrier2013}, and decrease the length of the accelerated neutral hydrogen tail behind the planet. Self-shielding means that the medium becomes optically thick in Ly$\alpha$, which prevents the UV photons from penetrating inside the deeper layers and interacting with hydrogen atoms. In the present study, we do not include either UV scattering or photoionization in atmospheres below $R_0$, which are protected from the UV radiation by the upper atmosphere. For the hot gas giants considered by \citet{Bourrier2013}, self-shielding is also important in the exosphere, but one should note that the hydrogen density in the upper atmosphere of a ``Hot Jupiter'' exceeds the upper atmosphere densities of the Kepler-11 ``super-Earths'' by approximately two orders of magnitude. For this reason, we do not include self-shielding effects into the DSMC upper atmosphere modeling presented in the current study.

\section{Ion production and pick-up escape}
\label{sec_pickup}

In this section, we present our estimate of the ion production rates for Kepler-11b--f.
These estimations are obtained in the same way as described in detail by \citet{Kislyakova2013}. Therefore, we only briefly summarize the main aspects in this study.

In the hydrogen-rich upper atmospheres of the five studied Kepler-11 planets, the H$^+$ ions of planetary origin are produced by charge-exchange reactions, photoionization and electron impact ionization processes. The intensity of the interactions and the number of ions produced vary from planet to planet and depend on upper planetary density and stellar wind parameters, which are functions of orbital distance for a particular star. Planetary obstacles and hydrogen coronae shapes also play a role. While we assume all Kepler-11 ``super-Earths'' are non-magnetized and have narrow magnetopauses/ionopauses close to the planets, the forms of the clouds are mostly defined by radiation pressure and charge-exchange and differ significantly from planet to planet (Fig.~\ref{f_clouds}). Strong radiation pressure accelerates neutral hydrogen atoms and moves them away from the planet, where they may undergo either charge-exchange or ionization by stellar XUV radiation or by stellar wind electrons easier.

Average ion production rates, $L_{\rm ion}$, which depend on coronal conditions related to heating efficiency are presented in Table~\ref{t_4}. Ion production and thermal loss rates are given in particles per second; mass loss rates are given in grams per second. Estimations of hydrogen thermal loss rates, $L_{\rm th}$, shown in the same table are taken from Table 3 of \citet{Lammer2013}. Thermal losses given in \citet{Lammer2013} present the calculation of losses from the whole atmosphere and start from the inner boundary of the simulation domain, which is well below Lagrange point L$_1$. In this paper, authors calculate how many particles can reach the L$_1$ point and are lost if the planet undergoes various levels of XUV heating. In the present paper, however, we consider a different loss mechanism, or so-called ion pickup. It is a nonthermal mechanism, when a particle does not obtain energy from thermal heating, but is accelerated and swept away from the planet by the stellar wind. An ion can be lost everywhere outside the magnetosphere/ionosphere of a planet. This mechanism has been observed to be effective for present-day Mars and Venus (see for example \citealp{Lundin2011}). Their atmospheres also do not extend beyond L$_1$ point. For this reason, the main atmospheres of Kepler-11 exoplanets also should not fill the whole Roche lobe to experience nonthermal ion pickup escape, and one can calculate pickup losses starting from magnetic boundary, as we do below.

As can be seen in Table~\ref{t_4}, the heating efficiency, $\eta$, also influences interactions between stellar winds and upper atmospheres. Higher values for $\eta$ result in higher ion production rates. As a consequence of a higher heating efficiency (40\% in considered case), the upper atmospheres expand more, which intensifies the stellar wind interaction processes. The corresponding stellar wind parameters (velocity, density, temperature) are discussed in Section~\ref{ssec_K11_sw} and can be found in Table~\ref{t_2}.

Ions produced beyond the planetary obstacle can be lost because of the ion pick-up process. Since these particles are no longer neutral, they are forced to follow the stellar wind plasma flow, detach from the planet, and finally escape from its gravity field. We consider the H$^+$ ions produced above the planetary obstacle, where the collisions between atmospheric particles can be neglected (Although we start at $Kn=0.1$, radiation pressure reshapes the exosphere and makes the hydrogen density drop off more rapidly on the starward side compared to the nondisturbed case). The presence of collisions would reduce the ion pickup escape and not change the main conclusion of the article, in which the nonthermal ion pickup is smaller compared to the thermal losses for exoplanets considered here. Taking into account the interplanetary magnetic field in the vicinity of the planets (see Table~\ref{t_2}), we can calculate the gyro radius for H$^+$:
\begin{equation}
	r_{\rm g} = \frac{m_{\rm H} v}{qB},
\end{equation}

\noindent where $m_{\rm H}$, $v$, and $q$ are the mass, velocity and electric charge of a hydrogen ion; $B$ is the stellar magnetic field near the planet; and $k_{\rm B}$ is the Boltzmann constant. We calculate two values of $r_{\rm g}$: $r_{\rm g}^I$ corresponds to the ``cold'' ions and $r_{\rm g}^{II}$ corresponds to ions that are accelerated by the electric fields in the stellar wind until they reach the speed of the stellar wind in the vicinity of the planet \citep{Lundin2011}. The  ``cold'' ion velocity is taken to equal the most probable Maxwellian speed $v_{\rm p} = \sqrt{2 k_{\rm B} T / m_{\rm H}} \approx 2.6$~km/s for $T=400$~K, which is the lowest temperature used in the simulations. As in \citet{Kislyakova2013}, we assume that an ion is lost from the planet if its gyro radius does not exceed $R_0$. This means that the magnetic field can significantly influence the trajectory of the particles. As can be seen in Table~\ref{t_4}, the gyro radius of H$^+$ ions is several orders of magnitude smaller than the planetary radius even in the most extreme case of Kepler-11f with $\eta$=40\%. Thus, we can assume that most newly produced ions are lost by Kepler-11 ``super-Earths'', so that ion production rates in Table~\ref{t_4} correspond to non-thermal ion pick-up loss rates.

\begin{table*}
\renewcommand{\baselinestretch}{1}
\caption{Ion production rates, thermal escape rates, and H$^+$ gyro radii for Kepler-11 planets. The values in lower lines are given in [g$\cdot$s$^{-1}$]. The value of $r_{g}^I$ corresponds to the speed of the H$^+$ of 2.6~km/s, and the value of $r_{g}^{II}$ to the stellar wind velocity in the vicinity of the planet. Thermal escape rates are taken from Table 3 of \citet{Lammer2013}.}
\begin{center}
\begin{tabular}{ccccccc}
\hline\hline
Exoplanet &  $L_{\rm ion}$, $\eta=$15\% [s$^{-1}$]  & $L_{\rm ion}$, $\eta=$40\% [s$^{-1}$]  & $r_{\rm g}^I$ [m] & $r_{\rm g}^{II}$ [m] &  $L_{\rm th}$, $\eta=$15\% [s$^{-1}$] & $L_{\rm th}$, $\eta=$40\%  [s$^{-1}$] \\\hline
Kepler-11b &   $\sim 7.0 \times 10^{30}$ &  $\sim 7.8 \times 10^{30}$  &  $\sim 7.0 \times 10^{1}$      & $\sim 1.1 \times 10^{4}$ & $\sim 7.0 \times 10^{31}$  &    $\sim 1.3 \times 10^{32}$ \\
 &   $\sim 1.17 \times 10^{7}$ &  $\sim 1.3 \times 10^{7}$  &       &  & $\sim 1.15 \times 10^{8}$  &    $\sim 2.0 \times 10^{8}$ \\

Kepler-11c &   $\sim 6.4 \times 10^{30}$ &  $\sim 8.2 \times 10^{30}$  &  $\sim 8.8 \times 10^{1}$      & $\sim 1.3 \times 10^{4}$ & $\sim 2.4 \times 10^{31}$  &    $\sim 7.3 \times 10^{31}$   \\
 &   $\sim 1.07 \times 10^{7}$ &  $\sim 1.37 \times 10^{7}$  &       &  & $\sim 4.0 \times 10^{7}$  &    $\sim 1.3 \times 10^{8}$ \\

Kepler-11d &   $\sim 8.8 \times 10^{30}$ &  $\sim 1.4 \times 10^{31}$  &  $\sim 2.2 \times 10^{2}$      & $\sim 3.6 \times 10^{4}$ & $\sim 6.0 \times 10^{31}$  &    $\sim 1.5 \times 10^{32}$  \\
 &   $\sim 1.47 \times 10^{7}$ &  $\sim 2.33 \times 10^{7}$  &       &  & $\sim 1.0 \times 10^{8}$  &    $\sim 2.5 \times 10^{8}$ \\

Kepler-11e &   $\sim 1.1 \times 10^{31}$ &  $\sim 2.0 \times 10^{31}$  &  $\sim 3.1 \times 10^{2}$      & $\sim 5.4 \times 10^{4}$ & $\sim 6.5 \times 10^{31}$  &    $\sim 1.5 \times 10^{32}$  \\
 &   $\sim 1.84 \times 10^{7}$ &  $\sim 3.34 \times 10^{7}$  &       &  & $\sim 1.1 \times 10^{8}$  &    $\sim 2.5 \times 10^{8}$ \\

Kepler-11f &   $\sim 3.6 \times 10^{31}$ &  $\sim 4.1 \times 10^{31}$  &  $\sim 5.5 \times 10^{2}$      & $\sim 8.9 \times 10^{4}$ & $\sim 2.5 \times 10^{32}$  &    $\sim 2.7 \times 10^{32}$  \\
 &   $\sim 6.0 \times 10^{7}$ &  $\sim 6.8 \times 10^{7}$  &       &  & $\sim 4.0 \times 10^{8}$  &    $\sim 4.5 \times 10^{8}$ \\
\hline
\end{tabular}
\end{center}
\normalsize
\label{t_4}
\end{table*}

\section{Conclusion}
\label{sec_conclusion}

In this study, we modeled the hydrogen coronae around the hydrogen-rich Kepler-11 ``super-Earths'' and studied the interactions between stellar wind plasma and their upper atmospheres. We estimated non-thermal ion pick-up losses and compared them to the thermal losses obtained from the hydrodynamic upper atmosphere model of \citet{Lammer2013}. We found that the newly produced planetary H$^+$ ions are mostly picked-up and swept away by the stellar wind in the Kepler-11 system, contributing to the total mass loss of the atmospheres of Kepler-11b--f. However, since we assumed that the planets have negligible magnetic moments, our estimated non-thermal ion pick-up rates represent upper limits. We found  that the non-thermal loss rates are approximately one order of magnitude smaller than the thermal losses estimated for the same planets by \citet{Lammer2013}. This ratio agrees well with results obtained by \citet{Kislyakova2013} for an Earth-type planet and a ``super-Earth'' in the habitable zone of a GJ436-like M-type host star. Since all Kepler-11 ``super-Earths'' are highly irradiated and in  a blow-off state (except of Kepler-11c, which still undergoes a very strong modified Jeans escape; see discussion in \citealp{Lammer2013}), it is  unlikely that non-thermal mass loss rates exceed thermal mass loss rates. Because of their closeness to their host star, stronger heating and interaction processes for Kepler-11b--f lead to higher thermal and ion pick-up escape rates than those obtained by \citet{Erkaev2013} and \cite{Kislyakova2013}. According to our estimations, stellar wind erosion of extended hydrogen coronae does not significantly enhance the total mass loss of massive hydrogen envelopes. Our results support the findings of \citet{Lammer2013} that many ``super-Earths'' can be expected to not completely lose their hydrogen envelopes by thermal and non-thermal atmospheric escape.

The detailed DSMC modeling of the hydrogen coronae of Kepler-11b--f also revealed some significant differences compared to the 1D-calculations of \citet{Lammer2013}. Although the net thermal escape rates are not affected, high radiation pressure and intense charge-exchange reshape the hydrogen cloud around the planet leading to strong asymmetry. This makes the density distribution not only dependent on radial but also on azimuthal and zenith coordinates and can affect temperature and heating distributions as well. These results show that 1D atmospheric modeling, which assumes a symmetric upper atmosphere, should be applied cautiously to highly irradiated exoplanets in the vicinities of their host stars.

\begin{acknowledgements}

K.G. Kislyakova, C.P. Johnstone, M.L. Khodachenko, H. Lammer, T. L\"{u}ftinger and M. G\"{u}del acknowledge the support by the FWF NFN project S116601-N16 ``Pathways to Habitability: From Disks to Active Stars, Planets and Life'', and the related FWF NFN subprojects, S116 604-N16 ``Radiation \& Wind Evolution from T Tauri Phase to ZAMS and Beyond'', S116 606-N16 ``Magnetospheric Electrodynamics of Exoplanets'', and S116607-N16 ``Particle/Radiative Interactions with Upper Atmospheres of Planetary Bodies Under Extreme Stellar Conditions''. T. L\"{u}ftinger acknowledges also the support by the FWF project P19962-N16.  K. G. Kislyakova, Yu. N. Kulikov, H. Lammer, and P. Odert thank also the Helmholtz Alliance project ``Planetary Evolution and Life''. P. Odert acknowledges support from the FWF project P22950-N16. The authors also acknowledge support from the EU FP7 project IMPEx (No.262863) and the EUROPLANET-RI projects, JRA3/EMDAF and the Na2 science WG5. N. V. Erkaev acknowledges support by the RFBR grant No 12-05-00152-a. Finally, the authors thank the International Space Science Institute (ISSI) in Bern, and the ISSI team ``Characterizing stellar- and exoplanetary environments''.  This research was conducted using resources provided by the Swedish National Infrastructure for Computing (SNIC) at the High Performance Computing Center North (HPC2N). The authors thank also the anonymous referee for his useful comments.

\end{acknowledgements}

\bibliographystyle{aa} 
\bibliography{references} 

\end{document}